\documentclass[aps,prl,superscriptaddress,twocolumn,floatfix,nofootinbib]{revtex4}
\usepackage{amssymb}
\usepackage{amsmath}
\usepackage{epsfig}
\usepackage{dcolumn}
\usepackage{rotating}
\usepackage{color}
\usepackage{hyperref}
\definecolor{rosso}{cmyk}{0,1,1,0.4}
\definecolor{rossos}{cmyk}{0,1,1,0.55}
\definecolor{rossoc}{cmyk}{0,1,1,0.2}
\definecolor{blu}{cmyk}{1,1,0,0.3}
\definecolor{blus}{cmyk}{1,1,0,0.6}
\definecolor{bluc}{cmyk}{1,1,0,0.1}
\definecolor{verde}{cmyk}{0.92,0,0.59,0.25}
\definecolor{verdec}{cmyk}{0.92,0,0.59,0.15}
\definecolor{verdes}{cmyk}{0.92,0,0.59,0.7}

\newcommand{\ba}{\begin{eqnarray}}
\newcommand{\ea}{\end{eqnarray}}
\newcommand{\be}{\begin{equation}}
\newcommand{\ee}{\end{equation}}
\newcommand{\bi}{\begin{itemize}}
\newcommand{\ei}{\end{itemize}}

\newcommand{\ga}{\gamma}

\newcommand{\da}{\delta}
\newcommand{\la}{\lambda}
\newcommand{\ka}{\kappa}

\newcommand{\sa}{\sigma}

\newcommand{\cO}{{\cal O}}

\newcommand{\p}{\partial}

\newcommand{\n}{\nabla}

\newcommand{\ra}{\rightarrow}

\newcommand{\LF}{\left(}
\newcommand{\RF}{\right)}
\newcommand{\LT}{\left[}
\newcommand{\RT}{\right]}

\newcommand{\2}{\frac{1}{2}}

\newcommand{\mx}{\mbox}
\newcommand{\mt}{\mathtt}

\newcommand{\with}{\mx{ with }}

\newcommand{\vs}{\vspace{0mm}\\}

\def\eq#1{(\ref{#1})}

\begin{document}

\title{Towards singularity and ghost free theories of gravity}

\author{Tirthabir Biswas}
\affiliation{Physics Department, Loyola University, Campus Box 92, New Orleans, LA 70118}
\author{Erik Gerwick}
\affiliation{II. Physikalisches Institut, Universit\"at G\"ottingen, Germany}
\author{Tomi Koivisto}
\affiliation{Institute for Theoretical Physics and Spinoza Institute, Postbus 80.195, 3508 TD
Utrecht, The Netherlands}
\affiliation{Institute of Theoretical Astrophysics, University of
  Oslo, P.O.\ Box 1029 Blindern, N-0315 Oslo, Norway}
\author{Anupam Mazumdar}
\affiliation{Physics Department, Lancaster University, Lancaster, LA1 4YB, United Kingdom}
\affiliation{Niels Bohr Institute, Blegdamsvej-17, Copenhagen-2100, Denmark}


\begin{abstract}
We present the most general covariant ghost-free  gravitational
action in a Minkowski vacuum. Apart from the much studied f(R) models, this
includes a large class of non-local actions with improved UV behavior, which
nevertheless recover Einstein's general relativity in the IR.
\end{abstract}


\maketitle


The theory of General Relativity (GR) has an ultraviolet (UV) problem which
is typically manifested in cosmological or black-hole type singularities.
Any resolution to this problem requires a theory which is
well behaved in the UV and reduces suitably to Einstein's gravity in
the infrared (IR)\footnote{In the light of current cosmic acceleration
observations, there have been efforts to modify gravity at large distances,
see \cite{woodward} for a review, but we do not discuss these models here.}. In
this letter, our aim is
to investigate whether the typical divergences at short distances
can be ameliorated in higher derivative covariant generalizations of GR.

Higher derivative theories of gravity are generally better behaved in the
UV and offer an improved chance to construct a singularity free
theory~\cite{bounce_strings}. Furthermore, Ref.~\cite{Stelle:1976gc}
demonstrated that fourth order theories of gravity are renormalizable,
but inevitably suffer from unphysical ghost states.  Therefore, before
we address the short-distance behavior of GR, we
first ennumerate the subset of all possible modifications to Einstein's gravity
which are guaranteed to be ghost-free.  To the best of our knowledge, a
systematic method for this is not presently available.
\vs
{\bf Generic quadratic action of gravity:} Let us start with
the most general covariant action of gravity. We immediately
realize that to understand both the asymptotic
behavior in the UV and the issue of ghosts, we require only
the graviton propagator.  In other words,  we look at metric
fluctuations around the Minkowski background
\be
g_{\mu\nu}=\eta_{\mu\nu}+h_{\mu\nu}\,,
\ee
and consider terms in the action that are quadratic in $h_{\mu\nu}$.
Since in the Minkowski background
$R_{\mu\nu\la\sa}$ vanishes, every appearance of the Riemann
tensor contributes an $\cO(h)$ term in the action. Hence,
we consider only terms that are products of at most  two
curvature terms, and higher ones simply do not play any role
in this analysis.  The most general relevant action  is of the form
\be\label{action1}
S=\int d^4x\sqrt{-g}\LT{R\over 2} + R_{\mu_1\nu_1\la_1\sa_1}
\cO_{\mu_2\nu_2\la_2\sa_2}^{\mu_1\nu_1\la_1\sa_1}
R^{\mu_2\nu_2\la_2\sa_2}\RT,
\ee
where $\cO$ is a differential operator containing covariant derivatives
and $\eta_{\mu\nu}$.  We note that 
if there is a differential operator
acting on the left Riemann tensor, one can always recast that
into the above form by integrating by parts. The most general action is
captured by 14 arbitrary functions, the $F_i$'s, which reduce to the 6 we display 
in eq.(\ref{a1}) upon repeated application of the Bianchi identities.

Our next task is to obtain the quadratic (in $h_{\mu\nu}$) free part of this action.
Since the curvature vanishes on the Minkowski background, the two
$h$ dependent terms must come from the two curvature terms present.  This
means the covariant derivatives take on their Minkowski values.
As is obvious, many of the terms simplify and combine to eventually produce the following action
\begin{alignat}{5}
& S_q   =  -  \int d^4x\Big[
\frac{1}{2}h_{\mu\nu} a(\Box) \Box h^{\mu\nu}
 +  h_{\mu}^{\sa} b(\Box)\p_{\sa}\p_{\nu}h^{\mu\nu}\label{lin_act} \\ \notag
 & +  h c(\Box)\p_{\mu}\p_{\nu}h^{\mu\nu}
+ \frac{1}{2}h d(\Box) \Box h
 +  h^{\la\sa} \frac{f(\Box)}{\Box}\p_{\sa}\p_{\la}\p_{\mu}\p_{\nu}h^{\mu\nu}\Big]\,.
\end{alignat}
The above can be thought of as a higher derivative generalization of the
action considered by van Nieuwenhuizen in Ref.~\cite{peter}. Here, we
have allowed $a,b,c,d$ and $f$ to be nonlinear functions of the
derivative operators that reduce in the appropriate limit to
the constants $a$, $b$, $c$ and $d$ of Ref.~\cite{peter}. The function
$f(\Box)$ appears only in higher derivative theories. In the
appendix (\ref{a_term}-\ref{f_term}) we have calculated the
contribution from the Einstein-Hilbert term and the higher derivative
modifications to the action in eq.(\ref{lin_act}). From the explicit
expressions we observe the following relationships:
\ba
a +b &=& 0 \label{Grelations1} \\
c +d &=& 0\\
b+c + f &=& 0 \,
\label{Grelations}
\ea
so that we are left with only two independent arbitrary functions.

The field equations can be derived straightforwardly to yield
\ba
a(\Box) \Box h_{\mu\nu}  & + &  b(\Box)\p_{\sa}( \p_{\nu}h_{\mu}^{\sa} + \p_{\mu}h_{\nu}^{\sa})
\nonumber \\ & + & c(\Box)(\eta_{\mu\nu}\p_{\rho}\p_{\sa}h^{\rho\sa}  +\p_{\mu}\p_{\nu}h)
 +   \eta_{\mu\nu}d(\Box)\Box h \nonumber \\
& + &   f(\Box) \Box^{-1} \p_{\sa}\p_{\la}\p_{\mu}\p_{\nu}h^{\la\sa}  =  -\ka\tau_{\mu\nu}\, .
\label{linearized-eqn}
\ea
While the matter sector obeys stress energy conservation,  the geometric part
is also conserved as a consequence of the generalized Bianchi identities:
\ba
-\ka\tau\nabla_\mu \tau^{\mu}_\nu = 0 & = & (a+b)\Box h^\mu_{\nu,\mu} + (c+d)\Box \p_\nu h
 \nonumber \\ & + & (b+c+f )h^{\alpha\beta}_{,\alpha\beta\nu}\, .
\ea
It is now clear why eqs.(\ref{Grelations1}-\ref{Grelations}) had to be satisfied.
\vs
{\bf Propagator and physical poles:} We are now well-equipped to calculate
the propagator. The above field equations can be written in the form
\be
\Pi_{\mu\nu}^{-1}{}^{\la\sa}h_{\la\sa}=\ka\tau_{\mu\nu}
\ee
where $\Pi_{\mu\nu}^{-1}{}^{\la\sa}$ is the inverse propagator. One obtains
the propagator using the spin projection operators $\{P^2,P_{s}^0,P_{w}^0,P_m^1\}$,
see Ref.~\cite{peter}.
They correspond to the spin-2, the two scalars, and the vector
projections, respectively.  These form a complete basis.
 Considering each sector separately and
taking into account the constraints in eq.(\ref{Grelations1}-\ref{Grelations}),
we eventually arrive at a rather simple result
\be\label{prop}
\Pi={P^2\over ak^2}+{P_{s}^0\over (a-3c)k^2}\, .
\ee
We note that the vector multiplet and the $w$-scalar have disappeared, and the
remaining $s$-scalar has decoupled from the tensorial structure. Further, since
we want to recover GR in the IR, we must have
\be
a(0)=c(0)=-b(0)=-d(0)  =1\ ,
\label{GRlimit}
\ee
 corresponding to the GR values. This also means that as $k^2\ra 0$ we have only the physical graviton propagator:
\ba
\lim_{k^2\ra 0}\Pi^{\mu\nu}{}_{\la\sa}&=&({P^2/ k^2})-({P_{s}^0/2 k^2})\,.
\ea
A few remarks are now in order: First, let us point out that although the $P_s$
residue at $k^2=0$ is negative, it is a benign ghost. In fact,  $P_s^0$ has
precisely the coefficient to cancel the unphysical longitudinal degrees of freedom
in the spin two part~\cite{peter}.
Thus, we conclude that provided eq.(\ref{GRlimit}) is satisfied, the $k^2=0$ pole just
describes the physical graviton state. Secondly, eq.(\ref{GRlimit}) essentially means
that $a$ and $c$ are non-singular analytic functions at $k^2=0$, and therefore cannot
contain non-local inverse derivative operators (such as $a(\Box)\sim {1/ \Box}$).

Let us next scrutinize some of the well known special cases: \\
{\bf $f(R)$ gravity:} they are a subclass of scalar-tensor theories
and are studied in great detail both in the context of early universe cosmology and
dark energy phenomenology.
Here, only the $F_1$ appears as a higher derivative contribution (see appendix).
According to our preceding arguments, we obtain the physical
states from the $R^2$ term. Since $a=1$, it is easy to see that only the
$s$-multiplet propagator is modified.  It now has two poles:
$\Pi\sim -1/2k^2(k^2-m^2)+\dots$. The  $k^2=0$ pole has, as usual,  the wrong sign
of the residue, while the second pole has the correct sign.  This represents an
additional scalar degree of freedom confirming the well known fact~\cite{solganik,chiba}.\\
{\bf Fourth order modification in $R_{\mu\nu}R^{\mu\nu}$:} They have also been considered in the literature. This corresponds to having an $F_2$ term (see appendix), which modifies the spin-2 propagator: $\Pi\sim P_2/k^2(k^2-m^2)+\dots$. The second pole necessarily has the wrong residue sign and corresponds to the well known Weyl ghost, Refs.~\cite{solganik,chiba}. In fact, this situation is quite typical: $f(R)$ type models can be ghost-free,  but they do not improve UV behavior, while modifications involving $ R_{\mu\nu\la\sa}$'s can improve the UV behavior~\cite{Stelle:1976gc} but typically contain the Weyl ghost!

To reconcile the two problems we now propose first to look at a special class of non-local models  with $f=0$ or equivalently $a=c$. The propagator then simplifies to:
\be\label{prop1}
\Pi^{\mu\nu}{}_{\la\sa}={1\over k^2a(-k^2)}\LF P^2-\2P_{s}^0\RF .
\ee
It is obvious that we are left with only a single arbitrary function $a(\Box)$,  since now $a=c=-b=-d$.
Most importantly, we now realize that as long as $a(\Box)$ has no zeroes, these theories contain
no new states as compared to GR, and only modify the graviton propagator. In particular, by
choosing $a(\Box)$ to be a suitable {\it entire function} we can indeed improve the UV behavior
of gravitons without introducing ghosts. This will be discussed below.
\vs
{\bf Singularity free gravity:} We now analyze the scalar potentials in these non-local
theories, focussing particularly on the short distance behavior. As is usual, we
solve the linearized modified Einstein's equations (\ref{linearized-eqn}) for a point
source:
\be
\tau_{\mu\nu}=\rho\da_\mu^0\da_\nu^0=m\da^3(\vec{r})\da_\mu^0\da_\nu^0\ .
\ee
Next, we compute the two potentials, $\Phi(r),~\Psi(r)$, corresponding to the metric
\be
ds^2=-(1+2\Phi)dt^2+(1-2\Psi)dx^2\ .
\label{newtonian-metric}
\ee
Due to the Bianchi identities \cite{Quandt:1990gc,Nesseris:2009jf}, we only need to solve the
trace and the $00$ component of eq.(\ref{linearized-eqn}). Since the Newtonian
potentials are static, the trace and 00 equation simplifies considerably to yield
\ba
(a -3c)\Box h+(4c -2a+f)\p_{\mu}\p_{\nu}h^{\mu\nu}
 &=& \ka \rho \nonumber \\
a\Box h_{00} + c\Box h  -c  \p_{\mu}\p_{\nu}h^{\mu\nu}& =&  -\ka\rho\,,
\label{trace-00}
\ea
which for the metric eq.(\ref{newtonian-metric}) simplify to
\ba
2(a -3c)[\n^2 \Phi-4\n^2 \Psi]
 &=& \ka \rho \nonumber \\
2(c-a)\n^2\Phi -4 c\n^2\Psi  & =&  -\ka\rho\,.
\label{trace-00a}
\ea
We are seeking functions $c(\Box)$ and $a(\Box)$, such that there are no ghosts and  no $1/r$ divergence at short distances.

For  $f=0$, the Newtonian potentials are solved easily:
\ba
4a(\n^2)\n^2\Phi =4 a(\n^2)\n^2\Psi  =  \ka\rho=\ka m\da^3(\vec{r}).
\label{newtonian}
\ea
Now, we know that in order to avoid the problem of ghosts,  $a(\Box)$ must
be an entire function.  Let us first illustrate the resolution of singularities by
considering the following functional dependence~\cite{bounce_strings}:
\be
a(\Box)=e^{-\Box/M^2}.
\label{exponential}
\ee
Such exponential kinetic operators appear frequently in string theory~\cite{strings}. In fact, quantum loops  in such stringy non-local scalar theories remain finite giving rise to interesting physics, such as linear Regge trajectories~\cite{regge} and thermal duality~\cite{thermal}.
 We note that there are a wide range of allowed possible energy scales for $M$, including
roughly the range between $\Lambda$ and $M_{pl}$.

Taking the Fourier components of eq.(\ref{newtonian}), in a straight forward manner one obtains
\be
\Phi(r) \sim {m\over M_p^2}\int d^3p\ {e^{i\vec{p}\vec{r}}\over p^2 a(-p^2)}= {4\pi m\over r M_p^2}\int {dp\over p} \frac{\sin{p\, r}}{a(-p^2)}.
\ee
We note that the $1/r$ divergent piece comes from the usual GR action, but now it
is ameliorated. For eq. (\ref{exponential}) we have
\be
\Phi(r) \sim  {m\over M_p^2 r}\int {dp\over p} e^{-p^2/M^2} \sin{(p\,r)}={\frac{m\pi}{2M_p^2\,r}}\mt{erf}\left(\frac{r M}{2} \right),
\ee
and the same for $\Psi(r)$.  We observe that as $r \ra \infty$, $\mt{erf}( r)\ra 1$, and
we recover the GR limit. On the other hand, as $r\ra 0$, $\mt{erf}(r)\ra r$, making the
Newtonian potential converge to a constant $ \sim mM/M_p^2$. Thus, although the matter source has
a delta function singularity, the Newtonian potentials remain finite! Further, provided $mM\ll M_p$, our linear approximation can be trusted all the way to $r\ra 0$.

Let us next verify the absence of singularities in the spin-2 sector.
This will allow us, for example, to derive a singularity free
quadrupole potential.  We enforce the Lorentz gauge as usual so
that the generalized field equations \eq{linearized-eqn} read
\ba \label{in_lor_g}
a\Box h_{\mu\nu}\; -\; \frac{f}{2} \p_{\mu}\p_{\nu}h
\; - \; \frac{c}{2}\eta_{\mu\nu}\Box h =  -\ka\tau_{\mu\nu}\,.
\ea
Again for $f=0$ we have a simple wave equation for the graviton $a(\Box) \Box \bar{h}_{\mu\nu}
= -\ka\tau_{\mu\nu}$. We invert Einstein's equations for $\bar{h}_{\mu\nu} $ to
obtain the Greens function,  $ \bar{G}_{\mu\nu} $, for a point-like
energy-momentum source.  In other words, we solve for
\ba
a(\Box) \Box \bar{G}_{\mu\nu}(x-y) = -\ka\tau_{\mu\nu}\delta^4(x-y),
\ea
Under the assumption of slowly varying
sources, one has
\ba
\bar{G}_{\mu\nu}(r) \sim   \frac{\ka}{r} \pi \mt{erf}\left[ \frac{r M}{2}\right] \tau_{\mu\nu} (r) \ ,
 \label{efun}
\ea
for $a(\Box)$ given in eq.\eq{exponential}.
We observe that in the limit $r \to 0$, the Greens function remains
singularity free.  The improved scaling takes effect roughly only
for $r < 1/M$.
\vs
{\bf Cosmological Singularities:}  The very general framework of this paper allows
us to consistently address the singularities in early universe cosmology.  As an
example, we note that a solution to eq.\eq{linearized-eqn} with
\be
h \sim \mt{diag}( 0,A\sin{\lambda t},A\sin{\lambda t},A\sin{\lambda t})\with A\ll 1
\label{cos_sol}
\ee
describes a Minkowski space-time with small oscillations \cite{Collins}.  This configuration
is singularity free.  Evaluating the field equations for eq.\eq{cos_sol} gives
the constraint $a(-\la^2)-3c(-\la^2) =0$.  Thus, our simple $f=0$ case is not sufficient and we require
an additional scalar degree of freedom in the s-multiplet.  Note that this
also explains why a solution such as eq.\eq{cos_sol} is absent in GR.  We
generalize to $f \ne 0$, but take special care to keep intact our results in
eq.(\ref{GRlimit}) and eq.\eq{newtonian}. The most general ghost-free parameterization
 for $a \ne c$  is
\be
c(\Box) \equiv {a(\Box) \over 3}\left[1+ 2\LF1-\frac{\Box}{m^2}\RF\tilde{c}(\Box)  \right],
\ee
where $\tilde{c}(\Box),a(\Box)$ are entire functions.  Note that $m^2\ra \infty$ and $\tilde{c}=1$ reproduces the $f=0$ limit.
We now find that eq.\eq{cos_sol} is a solution to the vacuum field equations with $\la = m$.
How the universe can grow in such models and also how the matter sector can influence the dynamics can possibly be addressed only with knowledge
of the full curvature terms.  We hope to investigate this in future work, but see Ref.~\cite{cyclic}
and~\cite{Brandenberger} for similar considerations.
\vs
{\bf Generality:}
How general are the above arguments leading to a lack of singularities? According to the
Weierstrass theorem any entire function is written as $a(\Box)=e^{-\ga(\Box)}$, where
$\ga(\Box)$ is an analytic function.  For a polynomial  $\ga(\Box)$ it is now easy to
see that if $\ga>0$ as $\Box\ra \infty$, the propagator is even more convergent than
the exponential case leading to non-singular UV behavior.
\vs
{\bf Conclusion:} We have shown that by allowing higher derivative non-local operators, we
may be able to render gravity singularity free
without introducing ghosts or any other pathologies around the Minkowski background. It
should be reasonably straight-forward to extend the analysis to DeSitter backgrounds by
including appropriate cosmological constants. In fact, requiring that the theory remains free
from ghosts around  different classical vacua may be a way to constrain the higher curvature
terms that didn't seem to play any role in our analysis. Other ways of constraining/determining
the higher curvature terms would be to look for additional symmetries or to try to extend Stelle's
renormalizability arguments to these non-local theories. Efforts in this direction have been
made~\cite{Modesto:2011kw}. Finally, it is known that one can obtain GR starting from the free
quadratic theory for $h_{\mu\nu}$ by consistently coupling to its own stress energy tensor. Similarly,
can one obtain {\it unique} consistent covariant extensions of the higher derivative quadratic
actions that we have considered? We leave these questions for future investigations.
\vs
{\bf Acknowledgements:} We would like to thank Alex Koshelev for pointing out some redundancies in the gravitational action. 
\nopagebreak
\begin{widetext}
\noindent
{\bf Appendix}\vspace{5mm}\\
The quadratic action in curvature reads
\ba
S_q&=&\int d^4x\sqrt{-g}[R F_1(\Box)R+R_{\mu\nu} F_2(\Box)R^{\mu\nu}   
+  R_{\mu\nu\la\sa} F_{3}(\Box)R^{\mu\nu\la\sa}
+ R F_4(\Box)\n_{\mu}\n_{\nu}\n_{\la}\n_{\sa}R^{\mu\nu\la\sa} 
\nonumber \\ & + &
R_{\mu}^{\nu_1\rho_1\sa_1} F_{5} (\Box)\n_{\rho_1}\n_{\sa_1}\n_{\nu_1}\n_{\nu}\n_{\rho}\n_{\sa}R^{\mu\nu\la\sa}
+
R^{\mu_1\nu_1\rho_1\sa_1} F_{6}(\Box)\n_{\rho_1}\n_{\sa_1}
\n_{\nu_1}\n_{\mu_1}\n_{\mu}\n_{\nu}\n_{\rho}\n_{\sa}R^{\mu\nu\la\sa}] \label{a1}
\ea
where we have used the Bianchi identities:
\be
\n_{\sa}R_{\mu\nu\la\rho}+\n_{\rho}R_{\mu\nu\sa\la}+\n_{\la}R_{\mu\nu\rho\sa}=0
\ee
to absorb all the other covariant terms into the above six. Further, in the $F_4$, $F_{5}$ and $F_{6}$ terms, one ends up with anticommutator of the covariant derivatives due to the anti-symmetric properties of the Reimann tensor, but these anticommutators produce a third curvature term, and therefore these terms 
are at least $\cO(h^3)$.  Thus, the coefficients of the free theory (\ref{lin_act}) in terms of the $F$'s are given by
\label{coefficients}
\begin{alignat}{5} \label{a_term}
a(\Box) =   1-\frac{1}{2} F_2 (\Box) \Box - 2 F_{3} (\Box) \Box\\
%
b(\Box) =  -1+\2 F_2 (\Box) \Box + 2 F_{3} (\Box) \Box
\\
%
c(\Box) = 1+ 2 F_1(\Box) \Box + \2 F_2 (\Box) \Box\\
%
d(\Box) = -1-2F_1(\Box) \Box  - \2 F_2 (\Box) \Box
\end{alignat}
\vspace*{-0.7cm}
\begin{alignat}{5}
\label{f_term}
f(\Box) =   -  2 F_1 (\Box) \Box - F_2 (\Box)\Box-  2  F_{3}(\Box) \Box . 
\end{alignat}

\end{widetext}


\end{document}